\documentclass[a4paper]{article}
\usepackage{listings}
\usepackage{xcolor}
\usepackage{multirow}
\usepackage{jheppub}  
\usepackage{comment}
\usepackage{dcolumn}
\usepackage[english]{babel}
\usepackage[utf8x]{inputenc}
\usepackage[T1]{fontenc}
\usepackage{palatino}
\usepackage{amsmath}
\usepackage{afterpage}
\usepackage{graphicx, subcaption}
\usepackage[colorinlistoftodos]{todonotes}
\usepackage[colorlinks=true]{hyperref}
\usepackage{makeidx}
\newcommand{\be}{\begin{equation}}  
\newcommand{\ee}{\end{equation}}  
\newcommand{\bea}{\begin{eqnarray}}  
\newcommand{\eea}{\end{eqnarray}}  
\makeindex
\begin{document}

\vspace*{1.2cm}

\thispagestyle{empty}
\begin{center}

{\LARGE \bf Effects of the initial-state geometry on D-meson production in pp 
and pPb collisions}

\par\vspace*{7mm}\par

{

\bigskip

\large \bf R. Terra$^1$, A. V. Giannini$^{2,3}$ and F. S. Navarra$^1$}

\bigskip

{\large \bf  E-Mail: richard.terra@usp.br\\ andregiannini@ufgd.edu.br\\ 
navarra@if.usp.br}

\bigskip

$^1$Instituto de F\'{\i}sica, Universidade de S\~{a}o Paulo,\\ 
Rua do Mat\~ao, 1371, CEP 05508-090,  S\~{a}o Paulo, SP, Brazil\\
$^2$ Faculdade de Ciências Exatas e Tecnologia,\\ Universidade Federal da 
Grande Dourados,\\ CEP 79804-970,  Dourados, MS, Brazil\\
$^3$ Departamento de F\'isica,\\ Universidade do Estado de Santa Catarina,\\ 
CEP 89219-710 Joinville, SC, Brazil

\bigskip

{\it Presented at the Workshop of Advances in QCD at the LHC and the EIC, CBPF, 
Rio de Janeiro, Brazil, November 9-15 2025}

\vspace*{15mm}

\end{center}
\vspace*{1mm}

\begin{abstract}

Going from lower to higher multiplicity events in proton-proton and 
proton-lead collisions, the data show a stronger than linear growth of 
the D-meson normalized yields. In this contribution we try to understand this 
behavior using a Monte Carlo event generator which implements the $k_T$-factorization formalism. We use different spatial distribution of matter in 
the proton and in the lead nucleus at the initial stage of these collisions. 
We find that, with all the tested spatial distributions, the model reproduces 
well the behavior seen in the data. We conclude that this observable is not 
appropriate for a detailed study of the spatial distribution of matter in the 
proton.  
\end{abstract}

 \section{Introduction}
 
In the last years the matter distribution inside the proton has become an 
important topic discussed by the international community of high-energy 
physics. After the measurement of the proton mass and scalar radius     
\cite{nature}, which revealed the extent of its gluon distribution, one 
important question remains: what is the shape of matter distribution in the 
interior of the proton? A possible answer to this question is the baryon junction (BJ).

The baryon junction is a configuration in which the valence quarks are 
located at the vertices of an ``Y'' like gluon string, also called string  
junction. The intermediate point, the Fermat point, is introduced in order 
to keep the baryon wave function gauge invariant, as discussed in           
\cite{veneziano}. An immediate consequence of the junction in the proton is 
that, in proton-proton collisions, the baryon number can be carried by sea    
quarks which arise from the string \cite{kharzeev-bj}. Since the baryons that 
can be formed from  sea quarks carry a small fraction of the baryon momentum, 
they can be stopped in the midrapidity region and can be responsible for the 
baryon number asymmetry on that region 
\cite{stopping-tribedy,stopping1,stopping2,stopping3,stopping4,
stopping5,bran22}. 
Also, lattice QCD simulations of the interaction between three static quarks 
observed the formation of flux tubes connected by the Fermat point 
\cite{lattice1,lattice2,lattice3}. Although there are reasons to believe that 
the baryon junction is real, this still needs to be confirmed.

In \cite{nosso,dissertacao} the authors  showed that the baryon junction is
responsible for enhancing the $J/\psi$ relative yields in high-multiplicity   
proton-proton (pp) and proton lead (pPb) collisions. They combined the color 
evaporation model  
and the two component model                
with BJ initial conditions. This relative yield  
of a meson is given by the yield of the meson per charged particle 
normalized by its minimum bias average value 
\cite{pPbjpsi,alice-psi7,alice-psi13,Alice-D-pp,Alice-D-pPb}. In general, 
the experimental data grow linearly in the low density region, but grow faster  
in the high-multiplicity region. For  other
theoretical approaches to these data, see  \cite{YLima,salazar,raju}.

In \cite{nosso_multidist} we calculated the multiplicity distribution of   
charged particles at midrapidity for pp and pPb collisions \cite{alice24}.    
We simulated the collisions with the help of the  MC-KLN event generator, 
which combines Glauber initial conditions \cite{Glauber-principal} and   
$k_T$-factorization \cite{paperrcbk} with KLN unintegrated 
gluon distributions (UGDs) \cite{kln1,kln2}. We used four types of nucleon 
structures: hard-sphere, Gaussian, analytical baryon junction (BJ1) and 
numerical baryon junction (BJ2), and studied their effects on the  
multiplicity distributions.

In this work we aim to extend the simulations performed in       
\cite{nosso_multidist} to the study of  D-meson relative yields at     
midrapidity in pp and pPb collisions \cite{Alice-D-pp,Alice-D-pPb}. 
We will address  D-meson production by convoluting the formalism    
with the $g \rightarrow D$ KKKS08 fragmentation function \cite{ffs08}.     
In the next section we present the simulation details. 
After that we show our  results and present our concluding remarks.

\section{Formalism}

\begin{figure}[h!]
\centering
\begin{tabular}{cccc}
    \includegraphics[scale=0.23]{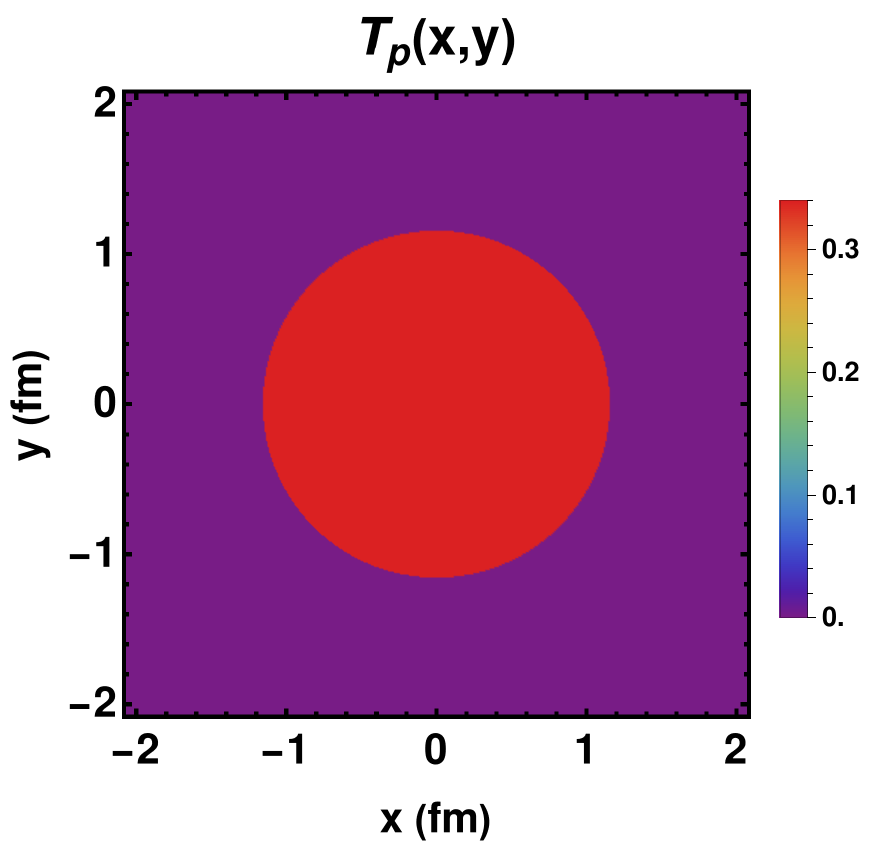}     & 
    \includegraphics[scale=0.23]{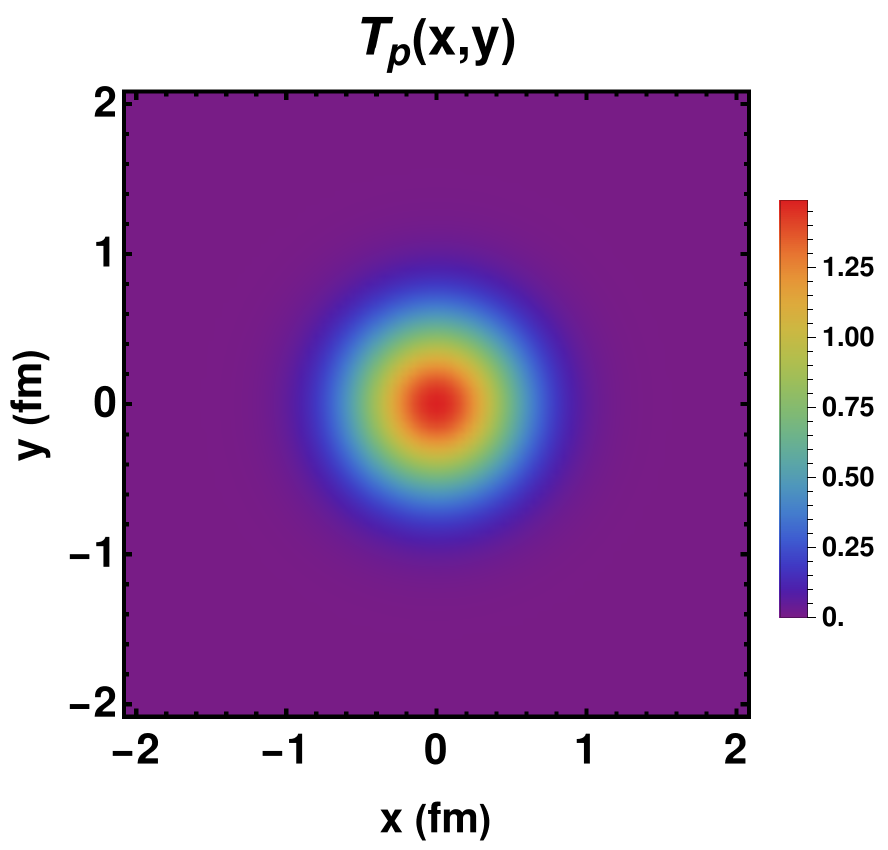} &
    \includegraphics[scale=0.23]{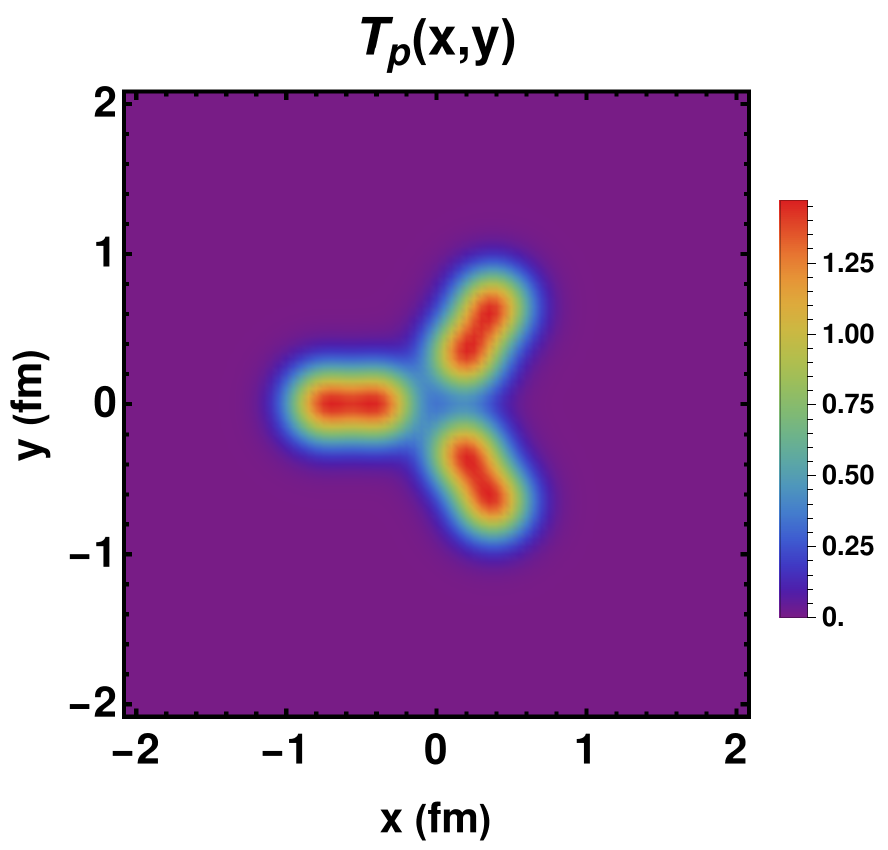}   & 
    \includegraphics[scale=0.23]{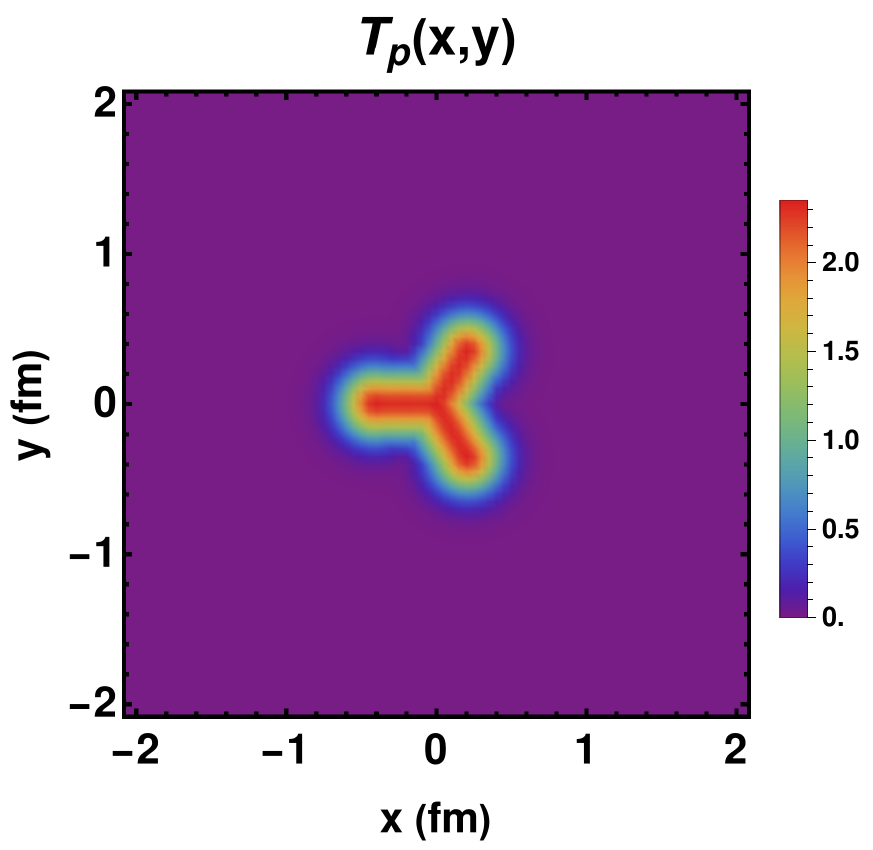} \\
  (a) &  (b) & (c) &  (d)\\
\end{tabular}
\caption{a) Hard-sphere, b) Gaussian nucleon,              
c) analytical baryon junction and d) numerical baryon junction 
thickness functions. Adapted from \cite{nosso_multidist}. }
\label{nucleons}
\end{figure}

Examples of the hard-sphere, Gaussian, BJ1 and BJ2 initial conditions for 
the nucleons used in the MC-KLN are shown in Fig. \ref{nucleons}. They are 
inserted in the code  through thickness functions (z-integrated densities) 
and overlap functions (probability per area of occurring nucleon-nucleon 
interactions). These functions are described in detail in           
\cite{nosso_multidist}. The BJ1 and BJ2 initial conditions are based on 
\cite{glazek-p, deb20} and \cite{schenke}, respectively.

\begin{figure}[h!]
\centering
\begin{tabular}{cccc}
    \includegraphics[scale=0.22]{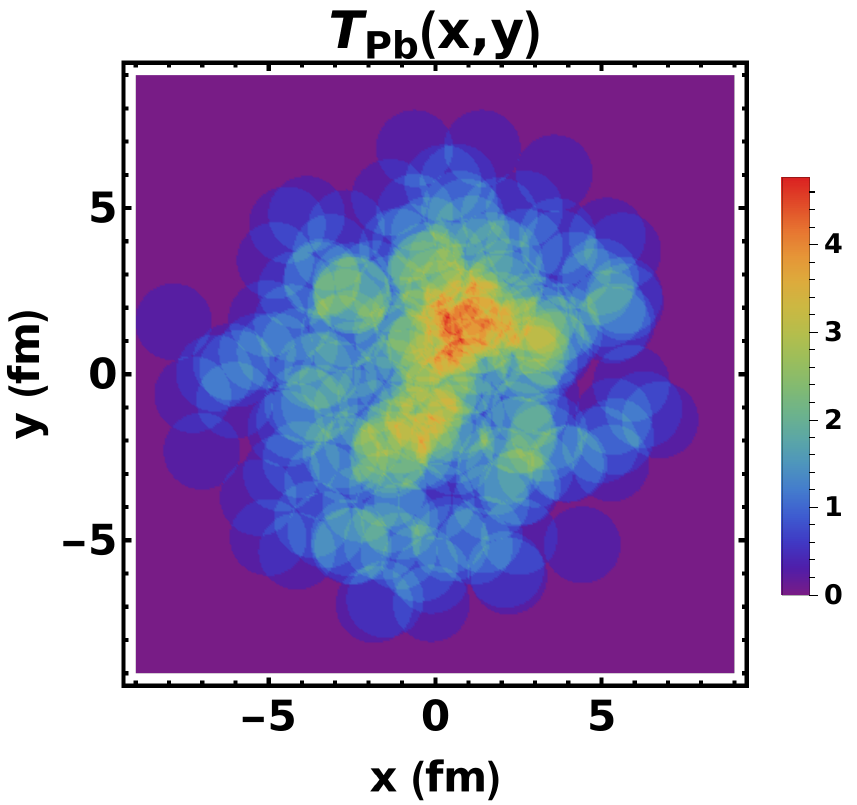}     & 
    \includegraphics[scale=0.22]{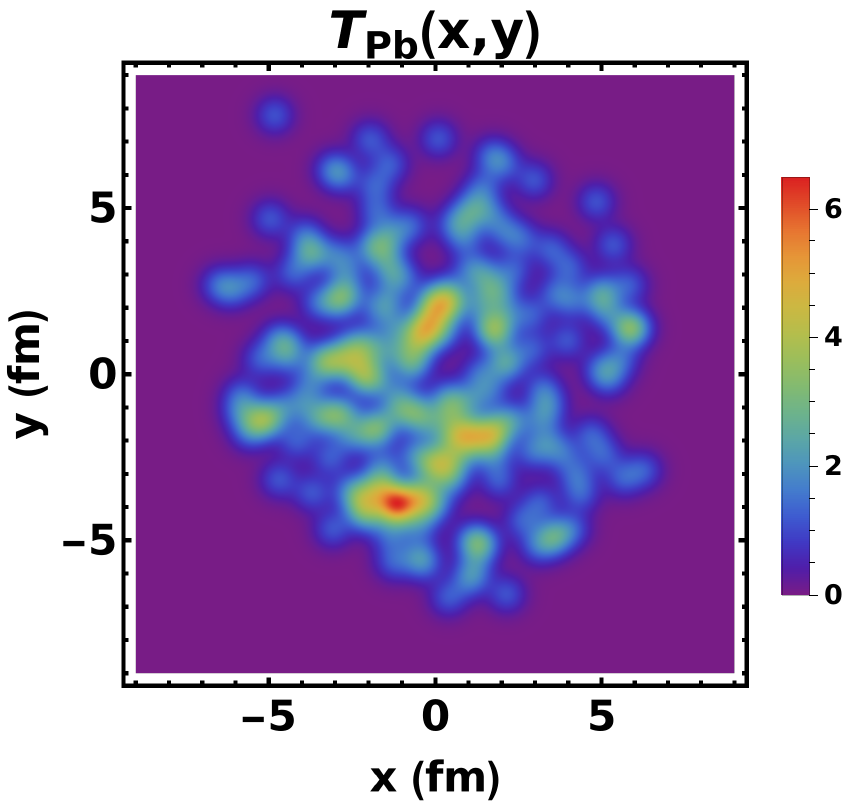} &
    \includegraphics[scale=0.22]{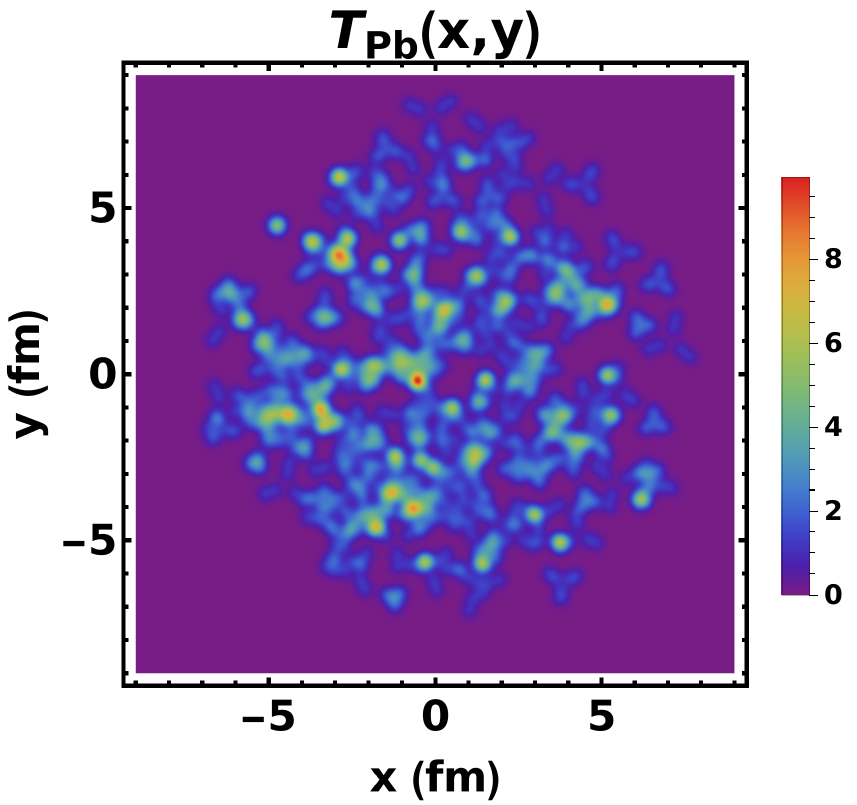}   & 
    \includegraphics[scale=0.22]{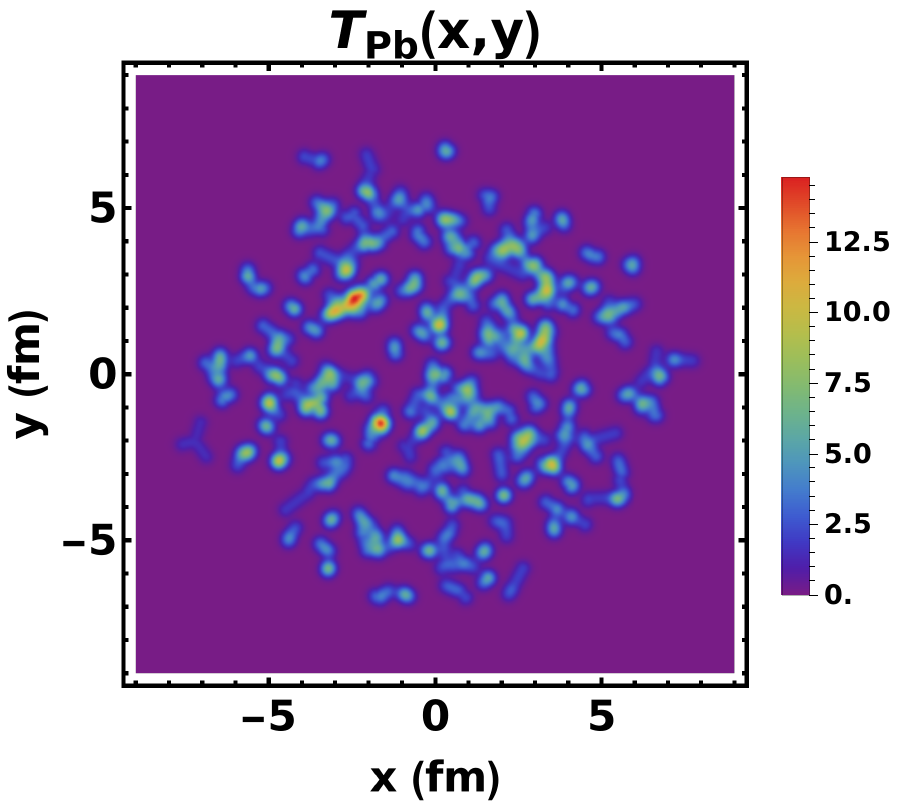} \\
  (a) &  (b) & (c) &  (d)\\
\end{tabular}
    \caption{Thickness functions of lead  composed by a) hard-sphere,       
b) Gaussian, c) analytical baryon junction and d) numerical baryon junction 
nucleons.}
\label{Pbs}
\end{figure}

For the lead initial condition, 208 nucleon centers are positioned in the space 
according the Woods-Saxon profile:
\begin{equation}
    \rho(r)=\frac{\rho_0}{1+\exp\big(\frac{r-R}{a}\big)} \, ,
    \label{nuclear charge density}
\end{equation}
Usual forms for this distribution are presented in \cite{Pbparam}. One example 
of thickness function of the lead  is shown in Fig. \ref{Pbs}.

Our event generator convolutes the projectile proton and the target proton  
(or lead) with the $k_T$-factorization single-inclusive gluon cross section 
\cite{paperrcbk}.
\begin{eqnarray}
\frac{d\sigma^{A+B \rightarrow g}}{dy \, d^2p_T \, d^2 \mathbf{r_{\perp}}} = 
\frac{2 \, K}{C_F \, p_T^2}
\int^{p_T} \frac{d^2k_T}{4} \int d^2b \text{ } \alpha_s \, 
\phi_p \bigg(\frac{|p_T+k_T|}{2},x_1;b \bigg) \, \phi_T 
\bigg(\frac{|p_T-k_T|}{2},x_2;R-b \bigg) \, .
    \label{ktfactorization}
\end{eqnarray}
In the above expression  $K$ is a multiplicative constant that accounts for    
higher order effects, $C_F$ is the color factor $C_F=(N_c^2-1)/2N_c$, $k_T$ is 
the intrinsic gluon transverse momentum and $p_T$ is the final state gluon      
transverse momentum. For the observables addressed by this work, $K$ has no  
effect on the data, since the results are normalized by the averaged value, 
and hence the constant $K$ is canceled. The $x$ values are calculated according to the $2 \rightarrow 1 $ kinematics:

\begin{equation}
x_{1,2}=\frac{p_T}{\sqrt{s}}e^{\pm y}  \, .
\end{equation}
$\phi$ is the KLN UGD \cite{kln1,kln2} given by 
\begin{equation}
\phi_{KLN}(\mathbf{k},x) = \frac{2C_F}{3\pi^2}\frac{(1-x)^4}{\alpha_s}  
\left \{ \begin{matrix} 1, & \text{, if } k\le Q_s \\ 
\big(Q_s/k\big)^2 & \text{, if } k> Q_s \end{matrix} \right. \, ,
    \label{KLN}
\end{equation}
calculated at the saturation scale \cite{Q2KLN}
\begin{equation}
Q_s^2(x,\mathbf{r_{\perp}})=T(\mathbf{r_{\perp}})\,\,               
\frac{2\text{ GeV}^{ 2}}{1.53 \text{ fm}^{-2}}\,\, 
\bigg(\frac{0.01}{x} \bigg)^{\bar{\lambda}} \, .
    \label{Q2}
\end{equation}
In the above expression, $\bar{\lambda}=0.23$ and $T(\mathbf{r}_\perp)$ is the 
projectile or the target initial  thickness, whose fluctuations will generate 
different final results. This is how the formalism is sensitive to variations 
in the spatial distribution of matter.

In order to reach the high-multiplicity region, fluctuations in the saturation  
scale (intrinsic fluctuations) are introduced in the formalism \cite{larry16}.  
These fluctuations are characterized by deviations of the saturation scale from 
its initial value $\left \langle Q_s^2 \right \rangle$ and are parametrized as a 
log-normal distribution: 
\begin{equation}
P(\ln(Q_s^2 / \left \langle Q_s^2 \right \rangle)) = \frac{1}{\sqrt{2\pi} 
\sigma} \exp \Bigg(-\frac{\ln^2(Q_s^2 / \left \langle 
Q_s^2 \right \rangle)}{2\sigma^2} \Bigg) \,.
    \label{dinamica}
\end{equation}
In the above expression the parameter $\sigma$ (the only free parameter in the  
formalism) determines the width of these fluctuations around the initial value.

To account for the charged particle multiplicity, we assume 
parton-hadron-duality, which implies that  charged particle production 
is proportional to  gluon production and the hadronization process does not
significantly change the final multiplicity. The simulations are done  using 
the $y \rightarrow \eta$ jacobian 
\begin{equation}
\frac{\cosh \eta}{\sqrt{\cosh^2\eta+m_\pi^2/k^2}} \,  ,
\end{equation}
in Eq.(\ref{ktfactorization}). This procedure allows us to calculate the 
pseudo-rapidity distribution $dN_{ch}/d \eta$ of charged particles in each 
collision.

For D-meson production, we convolute Eq. (\ref{ktfactorization}) with the 
$g \rightarrow D$ fragmentation functions given by the KKKS08 parametrization 
and taken at the factorization scale $\mu = m_D =1.8 \text{ GeV}$. In this way 
we can calculate the D-meson rapidity distribution, $dN_D/dy$, for each 
collision.

\section{Results} 

\begin{figure}[h!]
\centering
\begin{tabular}{cc}
    \includegraphics[scale=0.29]{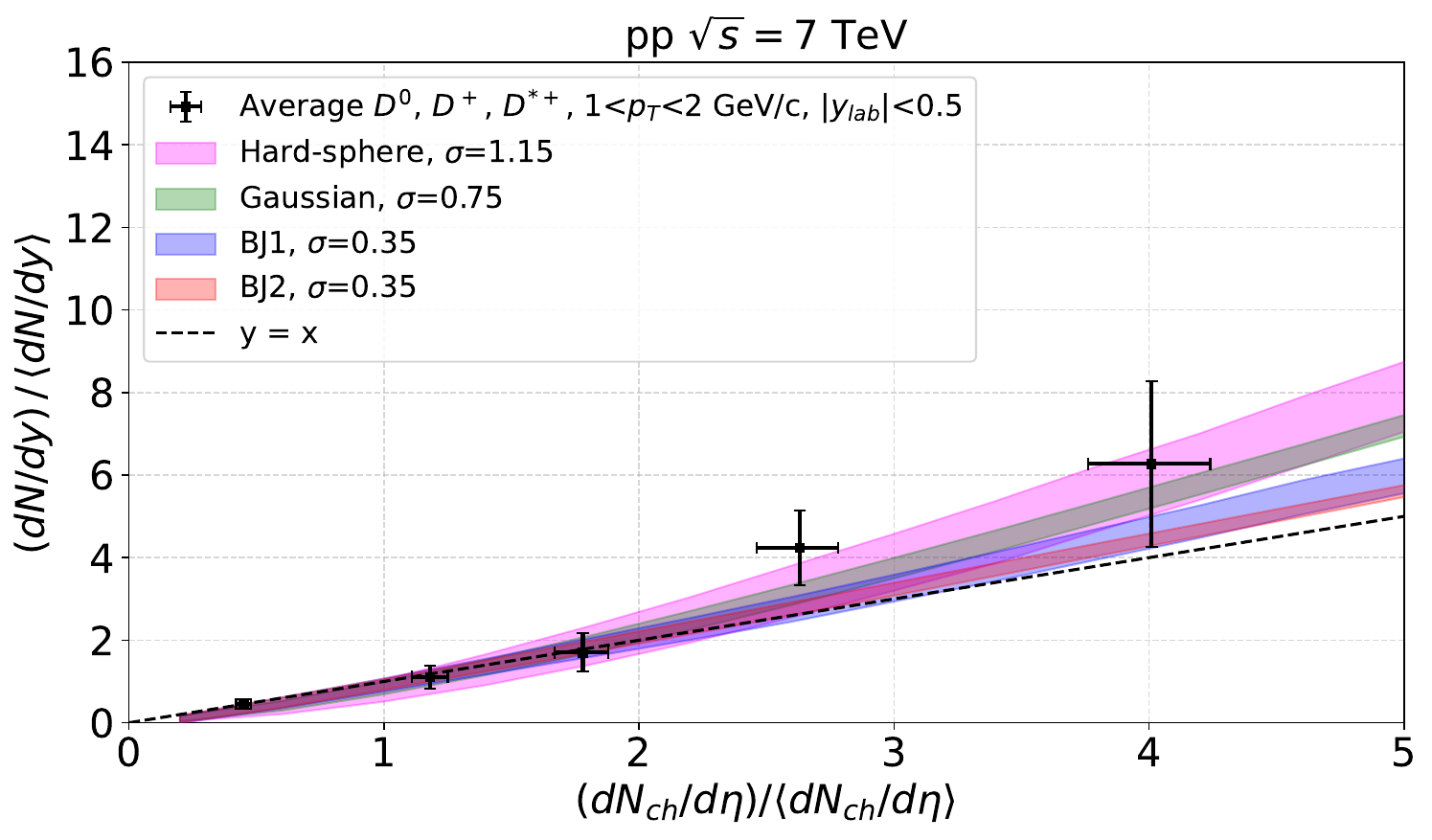}  & 
    \includegraphics[scale=0.29]{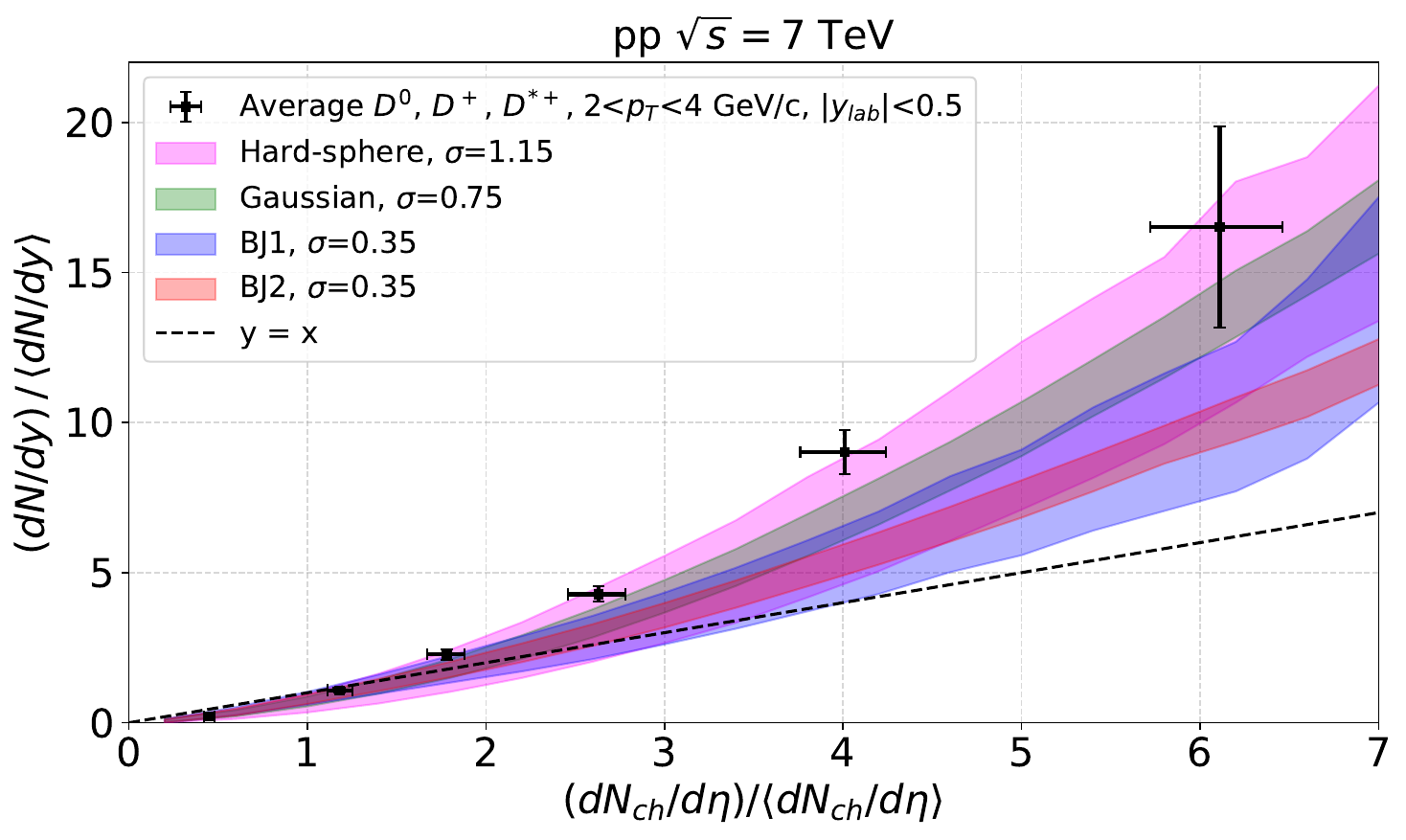}\\
    (a) &  (b) \\
\end{tabular}                                                    
\caption{Average $D^0$, $D^+$ and $D^{*+}$ relative yields in pp collisions in  
the range a) $1<p_T<2 \text{ GeV}$ and b) $2<p_T<4 \text{ GeV}$. The 
experimental data are from \cite{Alice-D-pp}.     }
\label{pp-results}
\end{figure}

\begin{figure}[h!]
\centering
\begin{tabular}{cc}
    \includegraphics[scale=0.29]{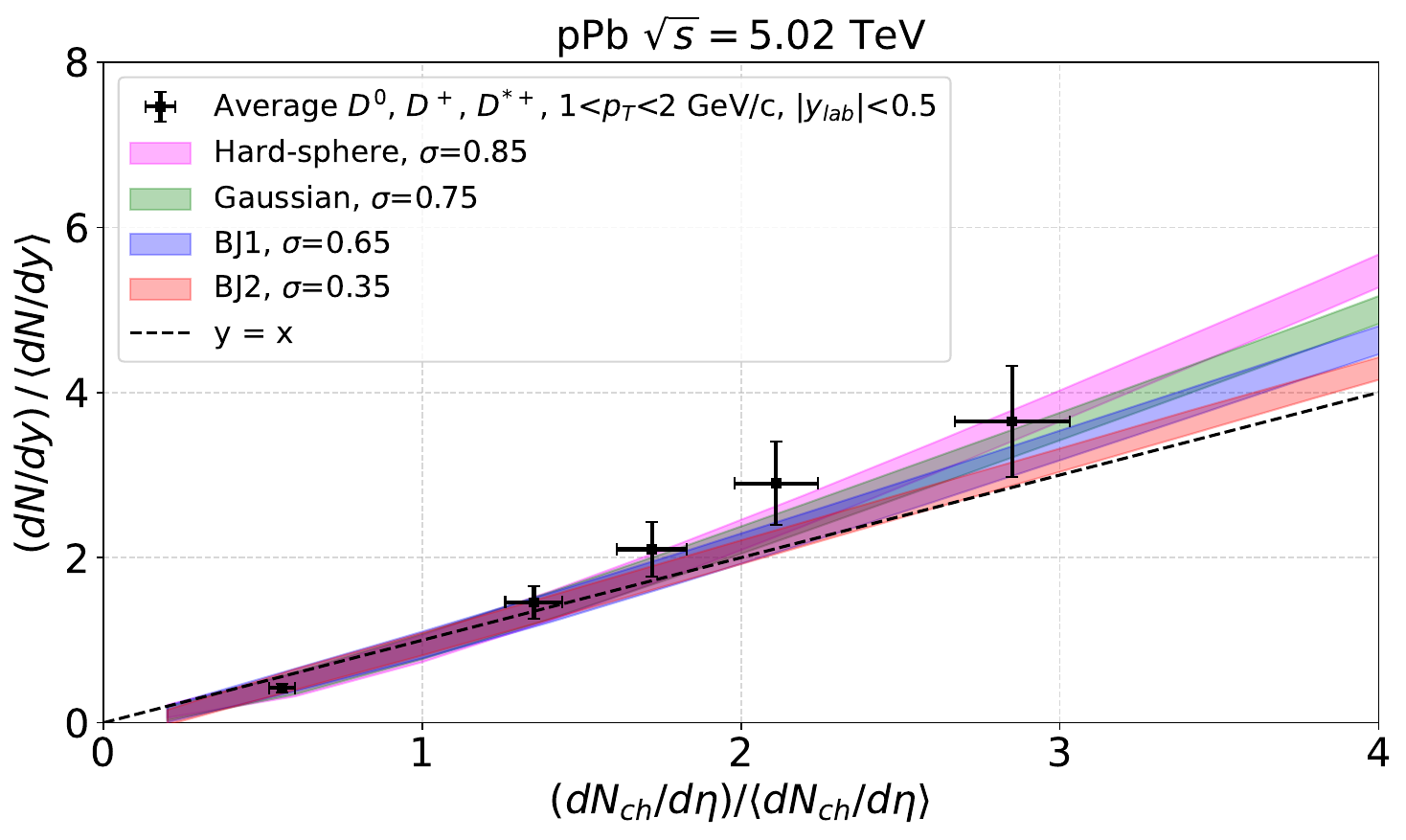}    & 
    \includegraphics[scale=0.29]{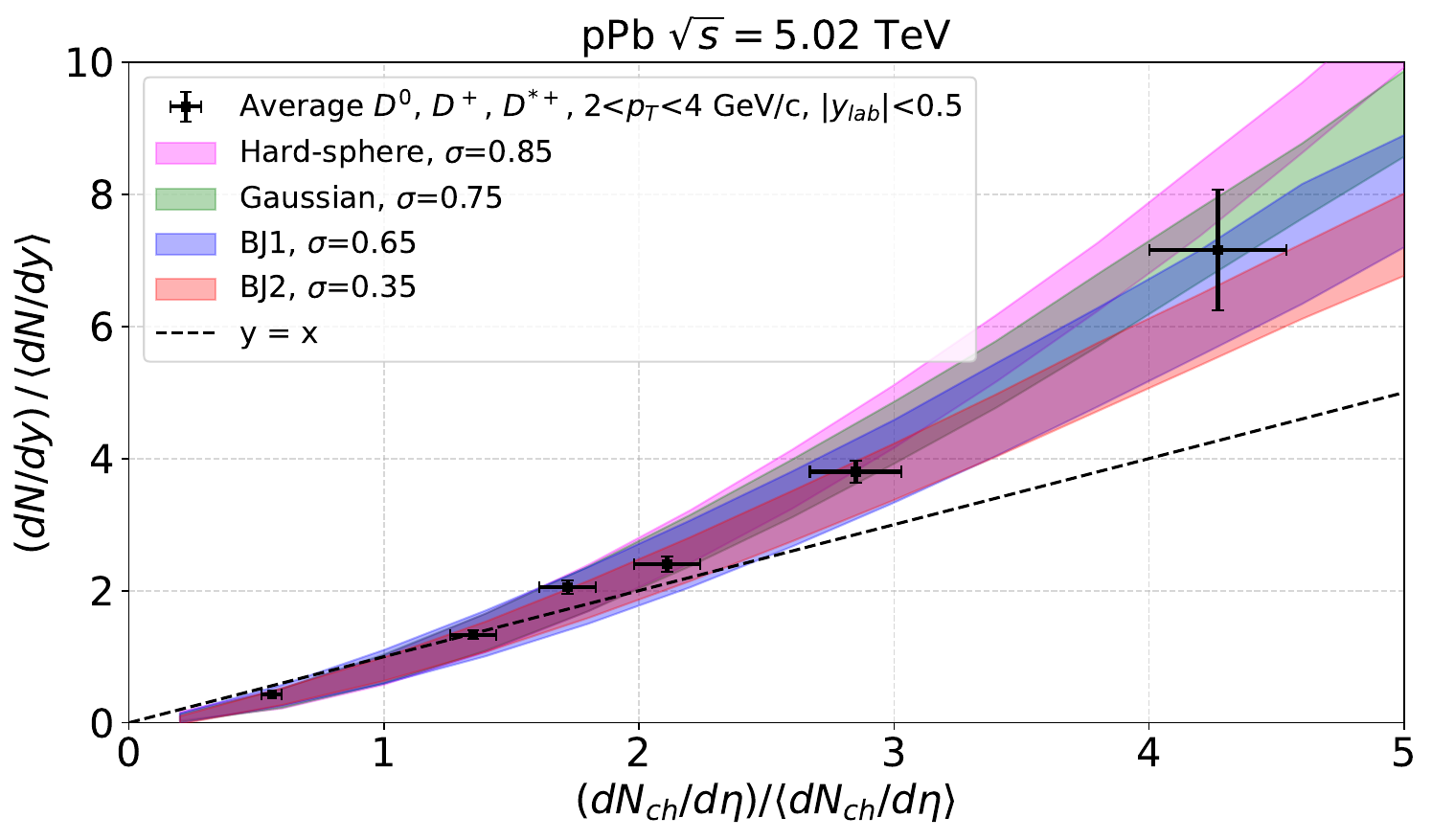} \\
  (a) &  (b) \\
\end{tabular}
\caption{Average $D^0$, $D^+$ and $D^{*+}$ relative yields for pPb         
collisions in the range a) $1<p_T<2 \text{ GeV}$ and b) $2<p_T<4 \text{ GeV}$. 
The experimental data are from \cite{Alice-D-pPb}.  }
\label{pPb-results}
\end{figure}

The results for pp and pPb collisions are shown in Fig.\ref{pp-results} and  
Fig.\ref{pPb-results}. In the figures we show the curves for each 
initial condition, the $\sigma$ value, the experimental data and a linear  
curve.  Because the data refer to  the midrapidity region, the $dN/d\eta$ 
and $dN_D/dy$  values are approximated by $dN/d\eta (\eta = 0) $ and           
$dN_D/dy(y = 0)$. The uncertainties are estimated by the standard deviations 
of the results on bins of size 0.4 in the x axis. 

In all the figures, the low-multiplicity results                           
($dN/d\eta/ \langle dN/d\eta \rangle \lesssim 2$) are compatible with each 
other and describe properly the experimental data. As the multiplicity 
increases, the theoretical curves start to split at                       
$dN/d\eta/ \langle dN/d\eta \rangle \approx 3$, except for  BJ1 and BJ2,   
which remain close to each other. 
Although the initial conditions yield different results    
on the high-multiplicity region ($dN/d\eta/ \langle dN/d\eta \rangle       
\gtrsim 4$), the experimental data on that region present large error bars and 
hence it is not possible to constrain the nucleon spatial configuration 
from these data.  
More statistics on future LHC runs could  
allow us to make stronger statements on the proton structure.

Although it is not possible to constrain the proton structure with these data,  
all the experimental data were explained by the formalism of gluon production 
and intrinsic fluctuations with Monte-Carlo simulations in the MC-KLN. 

\section{Conclusions}

In this work we used the MC-KLN event generator to simulate gluon production 
in pp and pPb collisions, for different nucleon geometrical configurations. 
We  obtained results for the D relative yields by using a gluon to         
D fragmentation function and by computing the charged particle through the 
parton-hadron duality. While all the initial conditions (hard-sphere, 
Gaussian, BJ1 and BJ2) could explain the  
data, it was not possible to constrain the proton structure because of the  
large errorbars in the high-multiplicity region. We expect to have a more  
conclusive constraint on the proton structure when the LHC delivers higher 
statistics data.   

\section*{Acknowledgements}

This study was financed, in part, by the São Paulo Research Foundation (FAPESP), 
Brasil. Process Number 2024/06652-4. This work was financed by the Brazilian    
funding agencies CNPq, FAPESP, CAPES, FAPERJ and by the INCT-FNA. This work has 
been done as a part of the Project INCT-Física Nuclear e Aplicações, Projeto 
No. 464898/2014-5.

The authors acknowledge the National Laboratory for Scientific Computing      
(LNCC/MCTI, Brazil), through the ambassador program (UFGD), subprojects FCNAE 
and SADFT for providing HPC resources of the SDumont supercomputer.


\end{document}